\documentclass[
 showpacs,preprintnumbers,
 amsmath,amssymb,
prb,
twocolumn ]{revtex4-1}

\usepackage{graphicx}
\usepackage{subfigure}
\usepackage{dcolumn}
\usepackage{bm}
\usepackage{bbm}
\usepackage{color}
\usepackage{esint}
\usepackage{lineno}

\begin{document}


\title{Majorana fermions in $s$-wave noncentrosymmetric superconductor with Dresselhaus (110) spin-orbit coupling}

\author{Jiabin You$^{1,}$}
\email{jiabinyou@gmail.com}

\author{C. H. Oh$^{1,}$}
\email{phyohch@nus.edu.sg}

\author{Vlatko Vedral$^{1,2,}$}
\email{phyvv@nus.edu.sg}

\affiliation{$^1$Centre for Quantum Technologies and Department of
Physics, National University of Singapore, 117543, Singapore\\
$^2$Department of Physics, University of Oxford, Clarendon
Laboratory, Oxford, OX1 3PU, United Kingdom}

\date{\today}

\begin{abstract}

The asymmetric spin-orbit interactions play a crucial role in
realizing topological phases in noncentrosymmetric superconductor
(NCS). We investigate the edge states and the vortex core states in
the $s$-wave NCS with Dresselhaus (110) spin-orbit coupling by both
numerical and analytical methods. In particular, we demonstrate that
there exists a novel semimetal phase characterized by the flat
Andreev bound states in the phase diagram of the $s$-wave
Dresselhaus NCS which supports the emergence of Majorana fermions.
The flat dispersion implies a peak in the density of states which
has a clear experimental signature in the tunneling conductance
measurements and the Majorana fermions proposed here should be
experimentally detectable.

\end{abstract}

\maketitle


\section{introduction}

Topological phase of condensed matter systems is a quantum many-body
state with nontrivial momentum or real space topology in the Hilbert
spaces
\cite{He.Volovik,PhysRevLett.49.405,PhysRevLett.61.2015,PhysRevLett.96.106802,JPSJ.77.031007,PhysRevLett.95.226801,PhysRevLett.95.146802,PhysRevLett.98.106803,PhysRevB.61.10267,PhysRevB.82.134521}.
Recent newly discovered topological superconductor (TSC) has spawned
considerable interests since this kind of topological phase supports
the emergence of Majorana fermion (MF)
\cite{Martin2004,PhysRevB.61.10267,PhysRevLett.100.096407,PhysRevB.82.134521,PhysRevLett.105.177002}
which is a promising candidate for the fault-tolerant topological
quantum computation (TQC) \cite{Kitaev20032}. There are several
proposals for hosting MFs in TSC, for example, chiral $p$-wave
superconductor \cite{PhysRevB.61.10267}, Cu-doped topological
insulator $\text{Bi}_{2}\text{Se}_{3}$
\cite{PhysRevLett.108.107005}, superconducting proximity devices
\cite{PhysRevLett.100.096407,Kitaev2001,Alicea2011,PhysRevB.81.125318,PhysRevB.86.161108}
and noncentrosymmetric superconductor (NCS)
\cite{PhysRevB.82.134521}. The signatures of MFs have also been
reported in the superconducting InSb nanowire \cite{Mourik2012},
$\text{Cu}_{x}\text{Bi}_{2}\text{Se}_{3}$
\cite{PhysRevLett.107.217001} and topological insulator Josephson
junction \cite{PhysRevLett.109.056803}. To obtain a readily
manipulated Majorana platform for TQC, more experimental
confirmations and theoretical proposals are therefore highly
desirable.

In this paper, we study the topological phase and Majorana fermion
at the edge and in the vortex core of the $s$-wave Dresselhaus (110)
spin-orbit (SO) coupled NCS. It is found that the asymmetric SO
interaction plays a crucial role in realizing topological phases in
the NCS. Although the Rashba SO coupled NCS has been previously
investigated \cite{ PhysRevB.82.134521,PhysRevLett.104.040502}, the
Dresselhaus (110) SO coupled NCS is relatively less discussed
theoretically \cite{PhysRevB.81.125318}. Interestingly, we find that
there is a novel semimetal phase in the Dresselhaus NCS, where the
energy gap closes in the whole region and different kinds of flat
Andreev bound states (ABSs) emerge. We demonstrate that these flat
ABSs support the emergence of MFs analytically and numerically. It
is known that the Chern number is not a well-defined topological
invariant in the gapless region, however, we find that the
topologically different semimetal phases in this gapless region can
still be distinguished by the Pfaffian invariant of the
particle-hole symmetric Hamiltonian.

Several authors have proposed the flat ABSs in the NCS
$\text{Li}_{2}\text{Pd}_{x}\text{Pt}_{3-x}\text{B}$ with high order
SO couplings \cite{PhysRevB.84.060504,PhysRevB.84.020501},
$d_{xy}$-wave superconductor, $p_{x}$-wave superconductor and
$d_{xy}+p$-wave superconductor \cite{PhysRevB.83.224511}. Instead,
our proposal for hosting the flat ABSs is an $s$-wave Dresselhaus
(110) SO coupled NCS in an in-plane magnetic field which is more
flexible than the previous proposals where one needs to apply a
magnetic field in the $y$ direction to the materials
\cite{PhysRevB.81.125318,PhysRevB.83.224511,Chris2012}. Our proposal
is experimentally more feasible. The flat dispersion implies a peak
in the density of states (DOS) which is clearly visible and has an
experimental signature in the tunneling conductance measurements
\cite{PhysRevLett.105.097002}. The zero-bias conductance peak has
been observed in recent experiments on the InSb nanowire
\cite{Mourik2012} and $\text{Cu}_{x}\text{Bi}_{2}\text{Se}_{3}$
\cite{PhysRevLett.107.217001} and argued to be due to the flat ABS.
Thus if the Majorana fermion exists in the Dresselhaus NCS, the flat
ABS and the zero-bias conductance peak in the DOS predicted here
should be detectable.

The paper is organized as follows. The model for $s$-wave NCS with
Dresselhaus (110) SO coupling is given in Sec. \ref{model}. The
phase diagrams and topological invariants of this model are
discussed in Sec. \ref{pdti}. The numerical and analytical solutions
to the Majorana fermions at the edge of the system are demonstrated
in Sec. \ref{MFatEdge}. The Majorana fermions in the vortex core of
the system are numerically shown in Sec. \ref{MFinvortex}. Finally,
we give a brief summary in Sec. \ref{summary}.

\section{model}\label{model}

We begin with modeling the Hamiltonian in a square lattice for the
two dimensional $s$-wave NCS with Dresselhaus (110) SO interaction
in an in-plane magnetic field, which is given by
$H=H_{\text{kin}}+H_{\text{Z}}+H_{\text{D}}^{110}+H_{s}$:
\begin{equation}
\label{eq1}
\begin{split}
H_{\text{kin}}=&-t\sum\limits_{is}\sum\limits_{\hat{\nu}=\hat{x},\hat{y}}(c_{i+\hat{\nu}s}^{\dag}c_{is}+c_{i-\hat{\nu}s}^{\dag}c_{is})-\mu\sum\limits_{is}c_{is}^{\dag}c_{is},\\
H_{\text{Z}}=&\sum\limits_{iss'}(\mathbf{V}\cdot\mathbf{\sigma})_{ss'}c_{is}^{\dag}c_{is'},\\
H_{\text{D}}^{110}=&-i\frac{\beta}{2}\sum\limits_{iss'}(\sigma_{z})_{ss'}(c_{i-\hat{x}s}^{\dag}c_{is'}-c_{i+\hat{x}s}^{\dag}c_{is'}),\\
H_{s}=&\sum\limits_{i}[\Delta_{s}c_{i\uparrow}^{\dag}c_{i\downarrow}^{\dag}+\text{H.c.}],\\
\end{split}
\end{equation}
where $c_{is}^{\dag}(c_{is})$ denotes the creation (annihilation)
operator of the electron with spin $s=(\uparrow,\downarrow)$ at site
$i=(i_{x},i_{y})$. $H_{\text{kin}}$ is the hopping term with hopping
amplitude $t$ and chemical potential $\mu$. $H_{\text{Z}}$ is the
Zeeman field induced by the in-plane magnetic field with components
$\mathbf{V}=(V_{x},V_{y},0)=\frac{g\mu_{B}}{2}(B_{x},B_{y},0)$.
$H_{\text{D}}^{110}$ is the Dresselhaus (110) SO coupling and
$H_{s}$ is the $s$-wave superconducting term with gap function
$\Delta_{s}$. We assume $t>0$ throughout this paper. In the momentum
space, the Hamiltonian is
$H=\frac{1}{2}\sum_{\mathbf{k}}\psi_{\mathbf{k}}^{\dag}\mathcal{H(\mathbf{k})}\psi_{\mathbf{k}}$
with
$\psi_{\mathbf{k}}^{\dag}=(c_{\mathbf{k}\uparrow}^{\dag},c_{\mathbf{k}\downarrow}^{\dag},c_{\mathbf{-k}\uparrow},c_{\mathbf{-k}\downarrow})$,
where
$c_{\mathbf{k}s}^{\dag}=(1/\sqrt{N})\sum_{\mathbf{l}}e^{i\mathbf{k}\cdot\mathbf{l}}c_{\mathbf{l}s}^{\dag}$,
$\mathbf{k}$ is the wave vector in the first Brillouin zone and the
Bogoliubov-de Gennes (BdG) Hamiltonian is
\begin{equation}
\label{eq2}
\begin{split}
\mathcal{H}(\mathbf{k})=\xi_{\mathbf{k}}\sigma_{z}+\beta\sin{k_{x}}\tau_{z}+V_{x}\sigma_{z}\tau_{x}+V_{y}\tau_{y}-\Delta_{s}\sigma_{y}\tau_{y},
\end{split}
\end{equation}
where $\xi_{\mathbf{k}}=-2t(\cos{k_{x}}+\cos{k_{y}})-\mu$,
$\mathbf{\sigma}$ and $\mathbf{\tau}$ are the Pauli matrices
operating on the particle-hole space and spin space, respectively.
The nontrivial topological order in the Dresselhaus NCS is
characterized by the existence of gapless edge state and Majorana
fermion. Below we shall demonstrate these features in the
Hamiltonian Eq. (\ref{eq1}).

\section{phase diagrams and topological invariants}\label{pdti}

\begin{figure}
\begin{center}
\includegraphics[width=8cm]{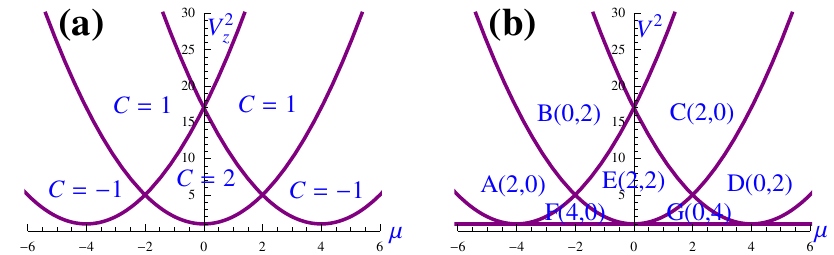}
\end{center}
\caption{(color online). The phase diagrams of $s$-wave (a) Rashba
and (b) Dresselhaus NCS. The parameters are $t=1$ and
$\Delta_{s}=1$. In (b), $V^{2}=V_{x}^{2}+V_{y}^{2}$. The Chern
number in different regions is indicated in (a). The number of
gap-closing points at $k_{x}=0$, $\nu_{1}$ and $k_{x}=\pi$,
$\nu_{2}$ in different regions are also shown as a pair $(\nu_{1},
\nu_{2})$ in (b).}\label{fig1}
\end{figure}

For comparison, we first briefly summarize the known results of the
$s$-wave Rashba NCS, in which the Dresselhaus (110) SO coupling
$H_{\text{D}}^{110}$ in the Hamiltonian Eq. (\ref{eq1}) is replaced
by the Rashba SO coupling
$H_{\text{R}}=-\frac{\alpha}{2}\sum_{i}[(c_{i-\hat{x}\downarrow}^{\dag}c_{i\uparrow}-c_{i+\hat{x}\downarrow}^{\dag}c_{i\uparrow})+i(c_{i-\hat{y}\downarrow}^{\dag}c_{i\uparrow}-c_{i+\hat{y}\downarrow}^{\dag}c_{i\uparrow})+\text{H.c.}]$
and the in-plane magnetic field is replaced by a perpendicular
magnetic field \cite{PhysRevB.82.134521}. As usual, we can use the
Chern number to characterize the nontrivial momentum space topology
of the Rashba NCS. The Chern number defined for the fully gapped
Hamiltonian is
$\mathcal{C}=\frac{1}{2\pi}\int_{T^2}dk_{x}dk_{y}\mathcal{F}(\mathbf{k})$,
where
$\mathcal{F}(\mathbf{k})=\epsilon^{ij}\partial_{k_{i}}A_{j}(\mathbf{k})$
is the strength of the gauge field
$A_{i}(\mathbf{k})=i\sum_{occ.}\langle\psi_{n}(\mathbf{k})|\partial_{k_{i}}\psi_{n}(\mathbf{k})\rangle$,
where $\psi_{n}(\mathbf{k})$ is the eigenstates of the Hamiltonian.
The integral is carried out in the first Brillouin zone and the
summation is carried out for the occupied states. As long as the
topological quantum transition does not happen, the Chern number
remains unchanged. Since the topological quantum transition happens
when the energy gap closes, the phase boundary can be depicted by
studying the gap-closing condition of the Hamiltonian. In the phase
diagram of the Rashba NCS as shown in the Fig. (\ref{fig1}a), we
find that the gap closes in some lines and the Chern number is
attached to each region of the phase diagram.

However, in the present case, we shall show that the phase diagram
of the Dresselhaus NCS has a gapless region that makes the Chern
number ill-defined. To see this, we diagonalize the BdG Hamiltonian
Eq. (\ref{eq2}) in the periodic boundary conditions of the $x$ and
$y$ directions, then the energy spectrum is
$E(\mathbf{k})=\pm\sqrt{\xi^{2}_{\mathbf{k}}+\mathcal{L}^{2}_{\mathbf{k}}+V^{2}+\Delta_{s}^{2}\pm2\sqrt{\xi^{2}_{\mathbf{k}}\mathcal{L}^{2}_{\mathbf{k}}+V^{2}(\xi^{2}_{\mathbf{k}}+\Delta_{s}^{2})}}
$, where $V=\sqrt{V_{x}^{2}+V_{y}^{2}}$ and
$\mathcal{L}_{\mathbf{k}}=\beta\sin{k_{x}}$. Therefore, we can find
that the energy gap closes at
$\xi^{2}_{\mathbf{k}}+\mathcal{L}^{2}_{\mathbf{k}}+V^{2}+\Delta_{s}^{2}=2\sqrt{\xi^{2}_{\mathbf{k}}\mathcal{L}^{2}_{\mathbf{k}}+V^{2}(\xi^{2}_{\mathbf{k}}+\Delta_{s}^{2})}$
which leads to the following gap-closing conditions:
$\xi^{2}_{\mathbf{k}}+\Delta_{s}^{2}=V^{2}$,
$\mathcal{L}_{\mathbf{k}}=0$. After some straightforward
calculations, we find that when $k_{x}=0$,
$(\mu+2t+2t\cos{k_{y}})^{2}+\Delta_{s}^{2}=V^{2}$; when $k_{x}=\pi$,
$(\mu-2t+2t\cos{k_{y}})^{2}+\Delta_{s}^{2}=V^{2}$. Finally, the gap
closes at $\{k_{x}=0,
\cos{k_{y}}=\frac{\pm\sqrt{V^{2}-\Delta_{s}^{2}}-\mu}{2t}-1\}$ or
$\{k_{x}=\pi,
\cos{k_{y}}=\frac{\pm\sqrt{V^{2}-\Delta_{s}^{2}}-\mu}{2t}+1\}$
subjected to $|\cos{k_{y}}|\leqslant1$. Therefore, we can find that
the gap closes in the regions from A to G as shown in the Fig.
(\ref{fig1}b). The number of gap-closing points at $k_{x}=0$,
$\nu_{1}$ and $k_{x}=\pi$, $\nu_{2}$ are also shown as a pair
$(\nu_{1},\nu_{2})$. Later we shall derive a relation between the
number of gap-closing points in the first Brillouin zone and the
topological invariant of the Hamiltonian. Interestingly, different
from the phase diagram of the Rashba NCS in the Fig. (\ref{fig1}a),
where the gap closes in some boundary lines and each gapped region
between them has a distinct Chern number, the phase diagram of the
Dresselhaus NCS has a gapless area from A to G as shown in the Fig.
(\ref{fig1}b), which means that the system is in the semimetal phase
in the whole region. Inside the gapless region, it is well known
that the Chern number is not well-defined. However, several other
topological invariants which are obtained from symmetry analysis of
the Hamiltonian can still be used to characterize the topologically
different semimetal phases in the gapless region. For the
Hamiltonian Eq. (\ref{eq2}), we enumerate several symmetries as
follow: (i) particle-hole symmetry,
$\Xi^{-1}\mathcal{H}(\mathbf{k})\Xi=-\mathcal{H}(-\mathbf{k})$; (ii)
partial particle-hole symmetry,
$\Xi^{-1}\mathcal{H}(k_{x},k_{y})\Xi=-\mathcal{H}(-k_{x},k_{y})$ and
(iii) chiral symmetry,
$\Sigma^{-1}\mathcal{H}(\mathbf{k})\Sigma=-\mathcal{H}(\mathbf{k})$,
where $\Xi=\sigma_{x}K$, $\Sigma=i\sigma_{y}\tau_{x}$ and $K$ is the
complex conjugation operator. We can define the Pfaffian invariant
\cite{PhysRevB.82.184525} for the particle-hole symmetric
Hamiltonian as
\begin{equation}
\label{pf}
\begin{split}
\mathcal{P}=\text{sgn}\left\{\frac{\text{Pf}[\mathcal{H}(\mathbf{K_{1}})\sigma_{x}]\text{Pf}[\mathcal{H}(\mathbf{K_{4}})\sigma_{x}]}{\text{Pf}[\mathcal{H}(\mathbf{K_{2}})\sigma_{x}]\text{Pf}[\mathcal{H}(\mathbf{K_{3}})\sigma_{x}]}\right\},
\end{split}
\end{equation}
where $\mathbf{K_{1}}=(0,0)$, $\mathbf{K_{2}}=(\pi,0)$,
$\mathbf{K_{3}}=(0,\pi)$ and $\mathbf{K_{4}}=(\pi,\pi)$ are the four
particle-hole symmetric momenta in the first Brillouin zone of the
square lattice. Similarly, the Pfaffian invariant
\cite{PhysRevLett.109.150408} for the partial particle-hole
symmetric system is
\begin{equation}
\label{pfky}
\begin{split}
\mathcal{P}(k_{y})=\text{sgn}\left\{\frac{\text{Pf}[\mathcal{H}(\pi,k_{y})\sigma_{x}]}{\text{Pf}[\mathcal{H}(0,k_{y})\sigma_{x}]}\right\}.
\end{split}
\end{equation}
For the chiral symmetry, if we take the basis where $\Sigma$ is
diagonal, $\Sigma=\text{diag}(i,i,-i,-i)$, then the Hamiltonian
becomes off-diagonal,
$\mathcal{H}(\mathbf{k})=\left(\begin{array}{*{20}c}
{\mathbf{0}} & {q(\mathbf{k})}\\
{q^{\dag}(\mathbf{k})} & {\mathbf{0}}\\
\end{array}\right)$. Using this $q(\mathbf{k})$, we can define the winding
number \cite{PhysRevB.83.224511} as
\begin{equation}
\label{wky}
\begin{split}
\mathcal{W}(k_{y})=\frac{i}{2\pi}\int_{-\pi}^{\pi}dk_{x}\text{tr}[q^{-1}(\mathbf{k})\partial_{k_{x}}q(\mathbf{k})].
\end{split}
\end{equation}
The Pfaffian invariant $\mathcal{P}$ can be used for identifying
topologically different semimetal phases of the Hamiltonian Eq.
(\ref{eq2}). It is easy to check that
$\mathcal{P}_{A}=\mathcal{P}_{B}=\mathcal{P}_{C}=\mathcal{P}_{D}=-1$
and $\mathcal{P}_{E}=\mathcal{P}_{F}=\mathcal{P}_{G}=1$ in the phase
diagram of the Dresselhaus NCS as shown in the Fig. (\ref{fig1}b).
Therefore, the semimetal phases in the region of A, B, C, D and the
region of E, F, G are topologically inequivalent. As for the other
two topological invariants $\mathcal{P}(k_{y})$ and
$\mathcal{W}(k_{y})$, below we shall show that they can be used to
determine the range of edge states in the edge Brillouin zone.

\section{Majorana Fermions at the edge of the system}\label{MFatEdge}

\begin{figure}
\begin{tabular}{c}
 \hspace{-0.02\textwidth}\includegraphics[width=8.5cm]{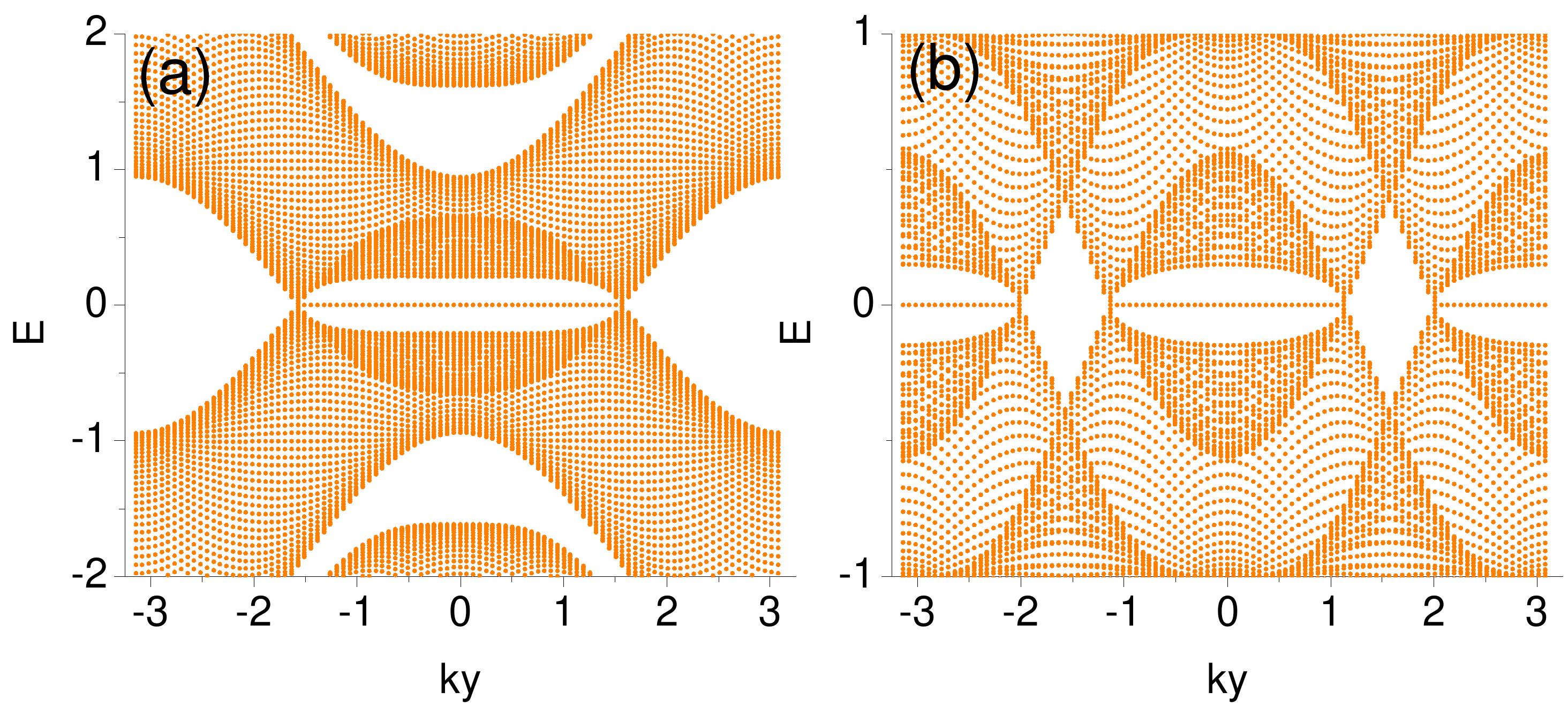}
\end{tabular}
\begin{tabular}{cc}
\includegraphics[width=4cm]{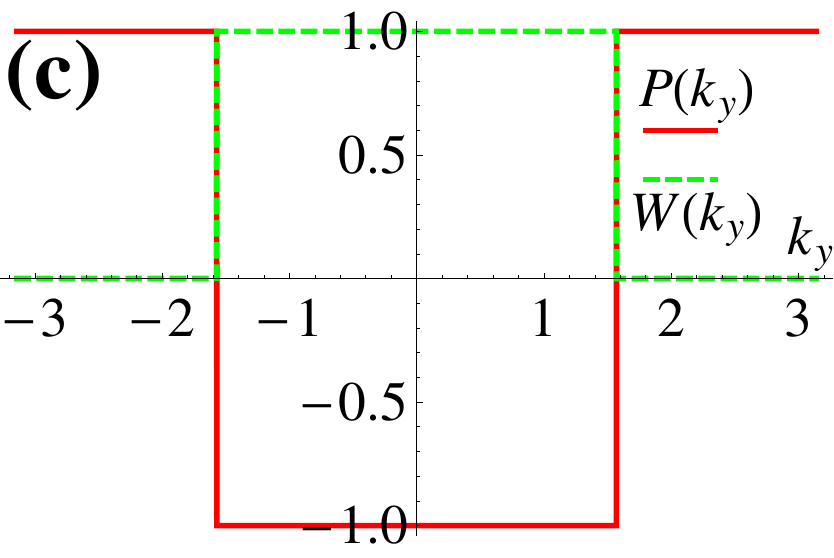} &
\includegraphics[width=4cm]{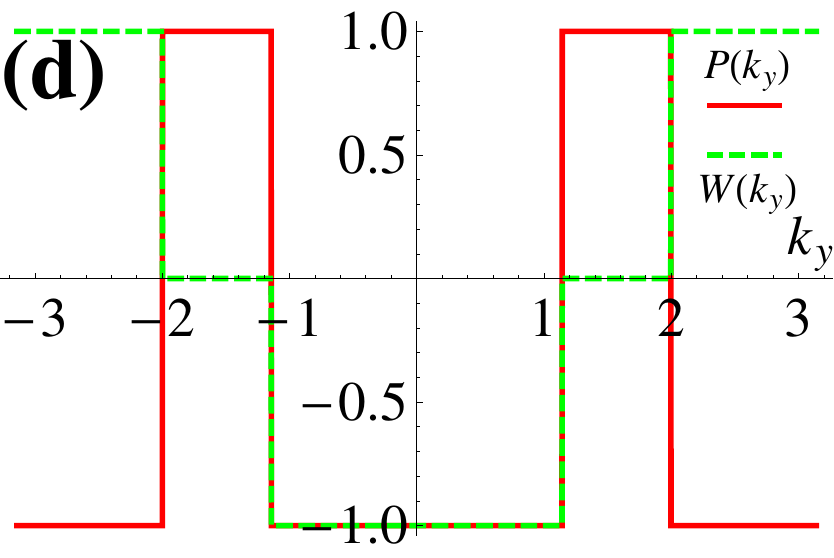} \\
\includegraphics[width=4cm]{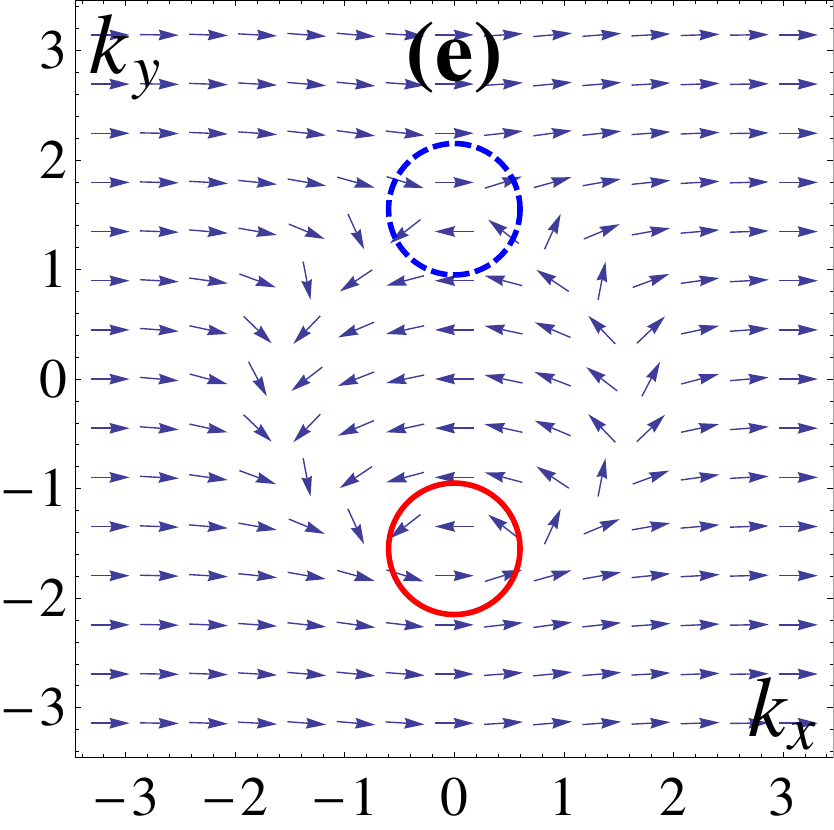} &
\includegraphics[width=4cm]{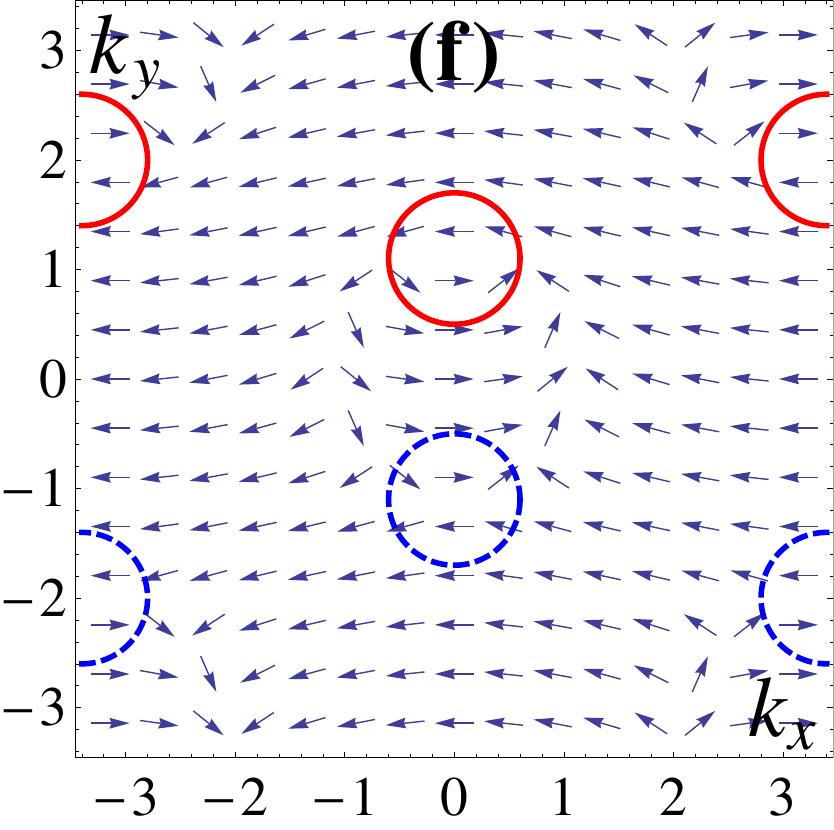} \\
\end{tabular}
\begin{tabular}{c}
\includegraphics[width=8cm,height=2cm]{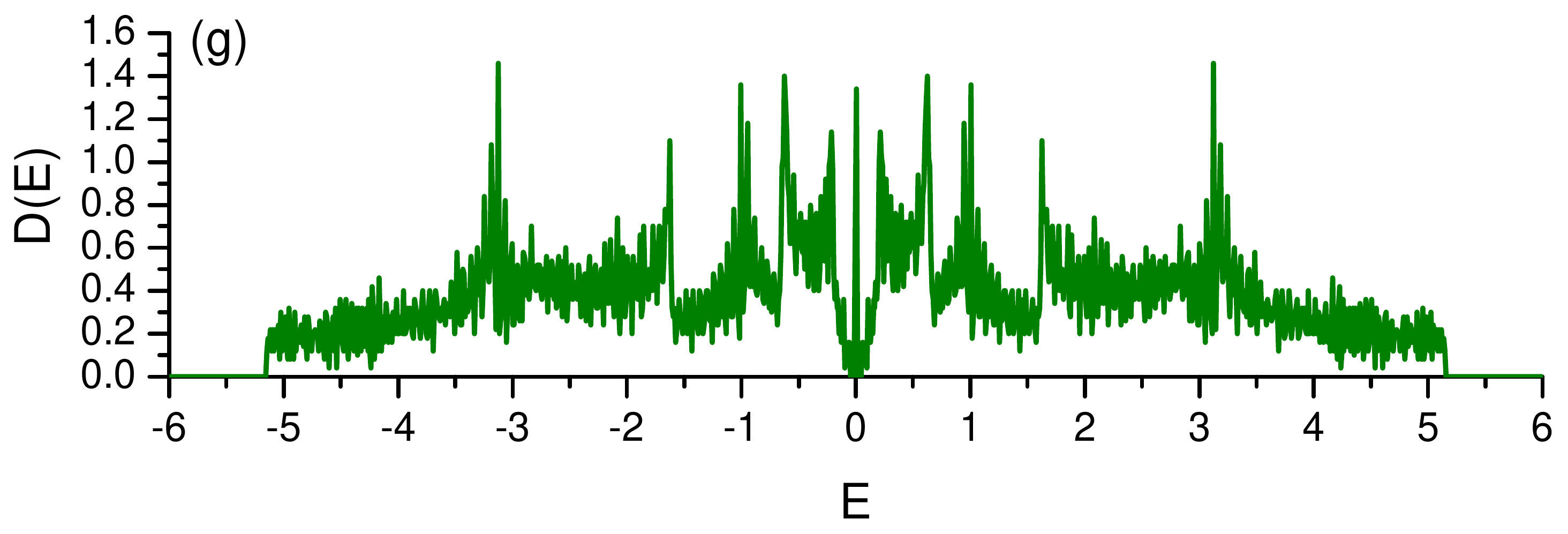} \\
\includegraphics[width=8cm,height=2cm]{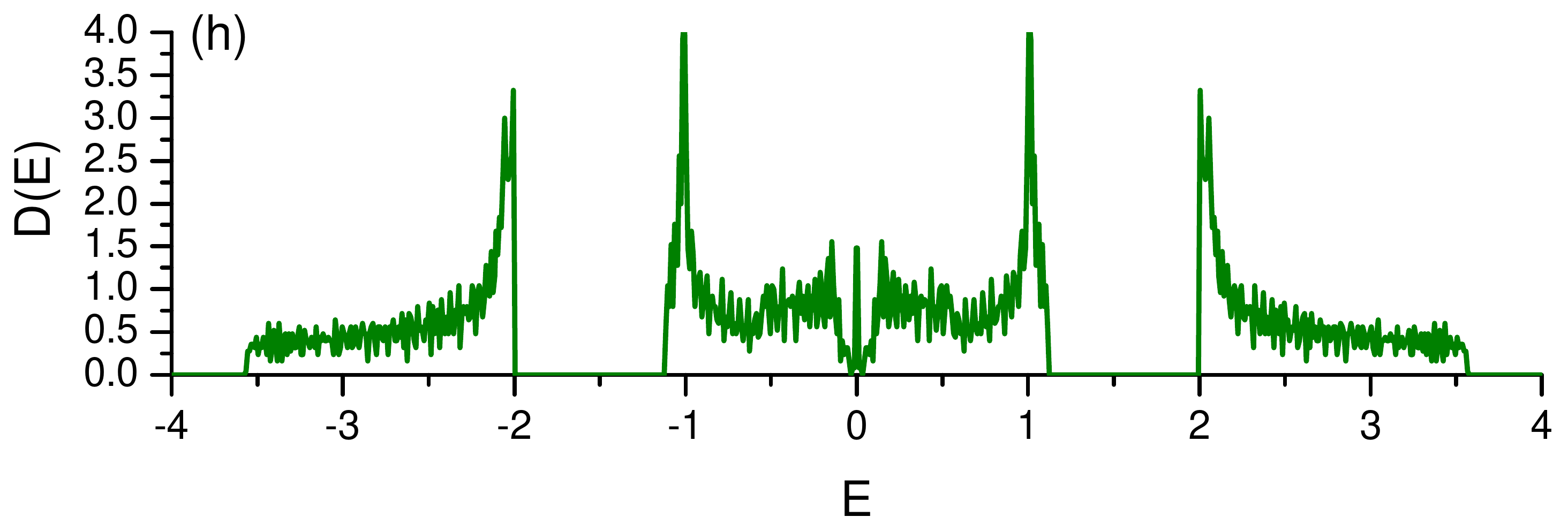} \\
\end{tabular}
\caption{(color online). (a) and (b) are the edge spectra of
$s$-wave NCS with Dresselhaus SO interaction. The open edges are at
$i_{x}=0$ and $i_{x}=50$, $k_{y}$ denotes the momentum in the $y$
direction and $k_{y}\in(-\pi,\pi]$. The parameters are $t=1$,
$\beta=1$, $\Delta_{s}=1$ and (a) $\mu=-4, V^{2}=5$, (b) $\mu=0,
V^{2}=9$, which correspond to region A and E, respectively. (c) and
(d) are the Pfaffian invariant Eq. (\ref{pfky}) and winding number
Eq. (\ref{wky}) for (a) and (b). (e) and (f) are the function
$z(\mathbf{k})$ for (a) and (b). The winding number of gap-closing
point enclosed by the red solid circle is $1$ and by the blue dashed
circle is $-1$, respectively. (g) and (h) are the DOS for (a) and
(b).}\label{fig2}
\end{figure}

To demonstrate the novel properties in the semimetal phase of the
Dresselhaus NCS, we study the Andreev bound states and Majorana
Fermions at the edge and in the vortex core of it. We now first turn
to study the ABSs of the Dresselhaus NCS. By setting the boundary
conditions of $x$ direction to be open and $y$ to be periodic, we
diagonalize the Hamiltonian Eq. (\ref{eq2}) in the cylindrical
symmetry and get the edge spectra of the Hamiltonian. Interestingly,
although the gap closes in the semimetal phase from region A to G as
shown in the Fig. (\ref{fig1}b), there exist dispersionless ABSs at
the edge of the system. The two topologically different semimetal
phases in the region A and E are depicted in the Fig. (\ref{fig2}a)
and (\ref{fig2}b), respectively. We would like to study the number
and range of the flat ABSs in these two different semimetal phases.
By the Pfaffian invariant Eq. (\ref{pfky}) or winding number Eq.
(\ref{wky}), the range where the flat ABSs exist in the edge
Brillouin zone can be exactly obtained as shown in the Fig.
(\ref{fig2}c) and (\ref{fig2}d). The number of flat ABSs is half of
the number of gap-closing points in the first Brillouin zone. From
the Hamiltonian in the chiral basis, we can see that the gap closes
when $\det q(\mathbf{k})=0$. In the complex plane of
$z(\mathbf{k})=\det q(\mathbf{k})/|\det q(\mathbf{k})|$, a winding
number can be assigned to each gap-closing point $\mathbf{k}_{0}$ as
$\mathcal{W}(\mathbf{k}_{0})=\frac{1}{2\pi
i}\ointctrclockwise_{\gamma}\frac{dz(\mathbf{k})}{z(\mathbf{k})-z(\mathbf{k}_{0})}$,
where $\gamma$ is a contour enclosing the gap-closing point. Due to
the particle-hole symmetry, we find that
$\mathcal{W}(\mathbf{k}_{0})=-\mathcal{W}(-\mathbf{k}_{0})$,
therefore, the gap-closing points with opposite winding number are
equal in number. The function $z(\mathbf{k})$ in the region A and E
are shown in the Fig. (\ref{fig2}e) and (\ref{fig2}f). As long as
the projection of opposite winding number gap-closing points does
not completely overlap in the edge Brillouin zone, there will be
flat ABSs connecting them \cite{PhysRevB.86.094512}. Therefore, the
number of flat ABSs is $\nu=(\nu_{1}+\nu_{2})/2$ and it is easy to
check that $\mathcal{P}$ is the parity of $\nu$,
$\mathcal{P}=(-1)^{\nu}$. The corresponding DOS of these two
different semimetal phases are shown in the Fig. (\ref{fig2}g) and
(\ref{fig2}h). We find that there is a peak at zero energy which is
clearly visible in the tunneling conductance measurements.
Therefore, all of these flat ABSs have clear experimental signature
in the tunneling conductance measurements and the MFs predicted at
the edge of the Dresselhaus NCS should be experimentally observable.
As for the robustness of the flat ABSs against disorder or impurity,
we can discuss it from the topological point of view. As long as the
disorder or impurity does not break the symmetries of Hamiltonian
Eq. (\ref{eq2}), these flat ABSs will be protected by the three
topological invariants mentioned above.

The existence of the edge states implies the nontrivial momentum
space topology in the Dresselhaus NCS so that the Majorana fermions
emerge at the edge of the system. In the following, we explicitly
calculate the Majorana zero modes at the edge of the Dresselhaus NCS
in the cylindrical symmetry. Let $x$ direction to be open and $y$ to
be periodic, then by setting $k_{x}\rightarrow-i\partial_{x}$, we
solve the Schr\"odinger equation of the Hamiltonian Eq. (\ref{eq2})
in the real space, $H(k_{x}\rightarrow-i\partial_{x},k_{y})\Psi=0$,
where
$\Psi=(u_{\uparrow},u_{\downarrow},v_{\uparrow},v_{\downarrow})^{T}$.
Due to the particle-hole symmetry in the Dresselhaus NCS, we have
$u_{\uparrow}=v_{\uparrow}^{*}$ and
$u_{\downarrow}=v_{\downarrow}^{*}$ at zero energy. Thus, we only
need to consider the upper block of the Hamiltonian Eq. (\ref{eq2}).
For simplicity, we consider the low energy theory at $k_{x}=0$, up
to the first order, we have
\begin{equation}
\label{eq4}
\begin{split}
(\varepsilon(k_{y})-i\beta\partial_{x})u_{\uparrow}+(V_{x}-iV_{y})u_{\downarrow}+\Delta_{s}u_{\downarrow}^{*}&=0,\\
(\varepsilon(k_{y})+i\beta\partial_{x})u_{\downarrow}+(V_{x}+iV_{y})u_{\uparrow}-\Delta_{s}u_{\uparrow}^{*}&=0,\\
\end{split}
\end{equation}
where $\varepsilon(k_{y})=-2t(1+\cos{k_{y}})-\mu$. Observing that
$u_{\uparrow}=\pm iu_{\downarrow}^{*}$, we obtain that when
$u_{\uparrow}=iu_{\downarrow}^{*}$, the solution is
$u_{\uparrow}(x)=c_{1}u_{\uparrow}^{1}(x)+c_{2}u_{\uparrow}^{2}(x)$,
where $c_{1}$ and $c_{2}$ are real numbers and
$u_{\uparrow}^{1}(x)=A_{1}e^{\lambda_{1}x}+A_{2}e^{\lambda_{2}x}$,
$u_{\uparrow}^{2}(x)=iB_{1}e^{\lambda_{1}x}+iB_{2}e^{\lambda_{2}x}$,
where
$\lambda_{1,2}=\frac{-\Delta_{s}\mp\sqrt{V^{2}-\varepsilon^{2}}}{\beta}$
and
$A_{1,2}=\frac{1}{2}(1\mp\frac{V_{x}-i(V_{y}+\varepsilon)}{\sqrt{V^{2}-\varepsilon^{2}}})$,
$B_{1,2}=\frac{1}{2}(1\pm\frac{V_{x}-i(V_{y}-\varepsilon)}{\sqrt{V^{2}-\varepsilon^{2}}})$;
when $u_{\uparrow}=-iu_{\downarrow}^{*}$, the solution is similar to
the case of $u_{\uparrow}=iu_{\downarrow}^{*}$. We consider the
Dresselhaus NCS in the positive $x$ plane with the edge located at
$x=0$. Let us assume $\Delta_{s}>0$ for simplicity, then from the
solutions of the Eq. (\ref{eq4}), the critical point for existing a
normalizable wavefunction under this boundary condition is
determined by $V^{2}-\varepsilon(k_{y})^{2}=\Delta_{s}^{2}$, which
is consistent with the gap-closing condition
$(\mu+2t+2t\cos{k_{y}})^{2}+\Delta_{s}^{2}=V^{2}$ at $k_{x}=0$. By
the same reason, the condition for normalizable wavefunctions is
consistent with the gap-closing condition
$(\mu-2t+2t\cos{k_{y}})^{2}+\Delta_{s}^{2}=V^{2}$ if we consider the
low energy theory at $k_{x}=\pi$. Therefore, the Majorana bound
state is
$(u_{\uparrow},iu_{\uparrow}^{*},u_{\uparrow}^{*},-iu_{\uparrow})^{T}$,
where $u_{\uparrow}$ is the solution of Eq. (\ref{eq4}).

\section{Majorana Fermions in the vortex core of the system}\label{MFinvortex}

\begin{figure}
\begin{tabular}{cc}
\includegraphics[width=4cm]{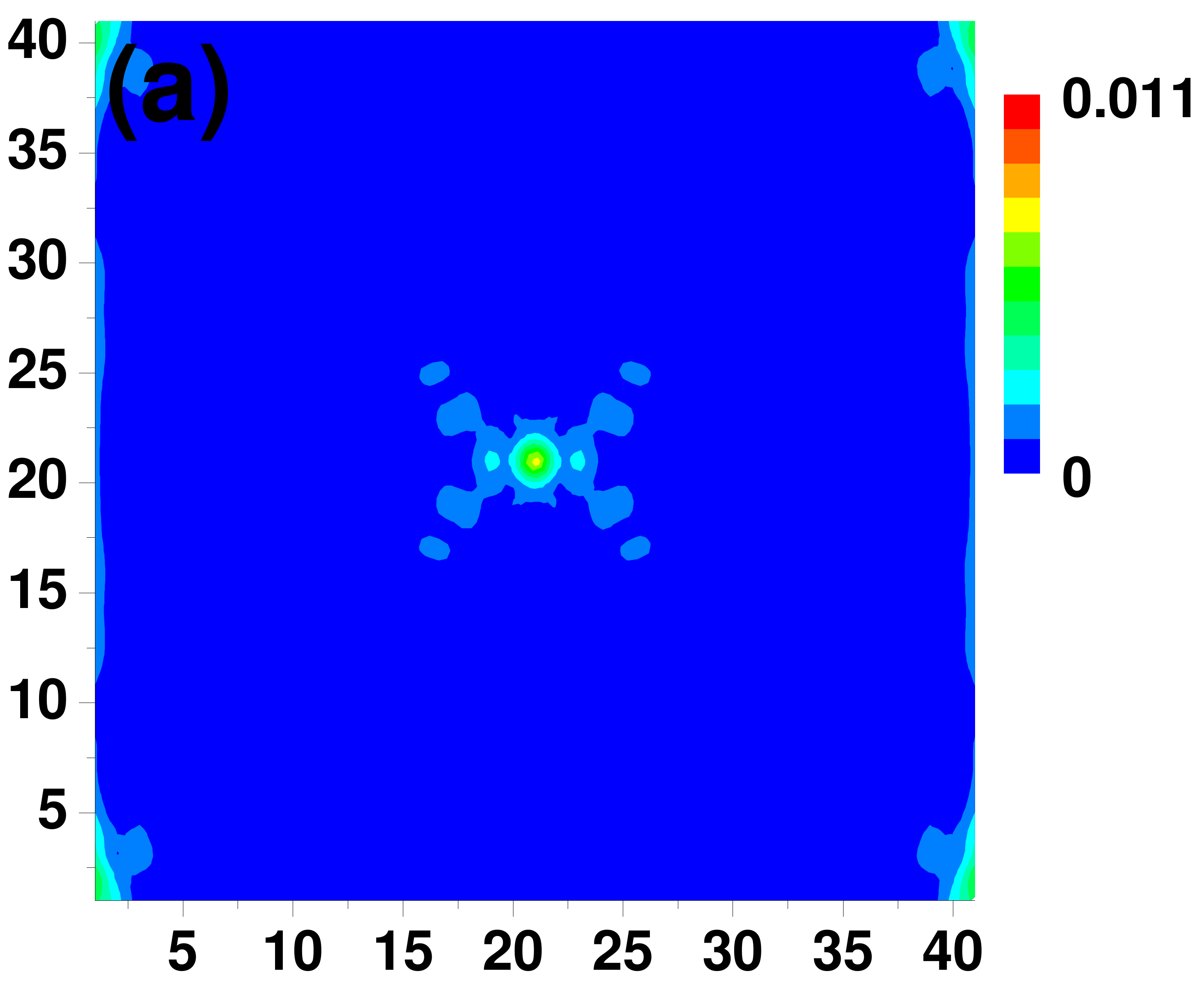} &
\includegraphics[width=4.1cm]{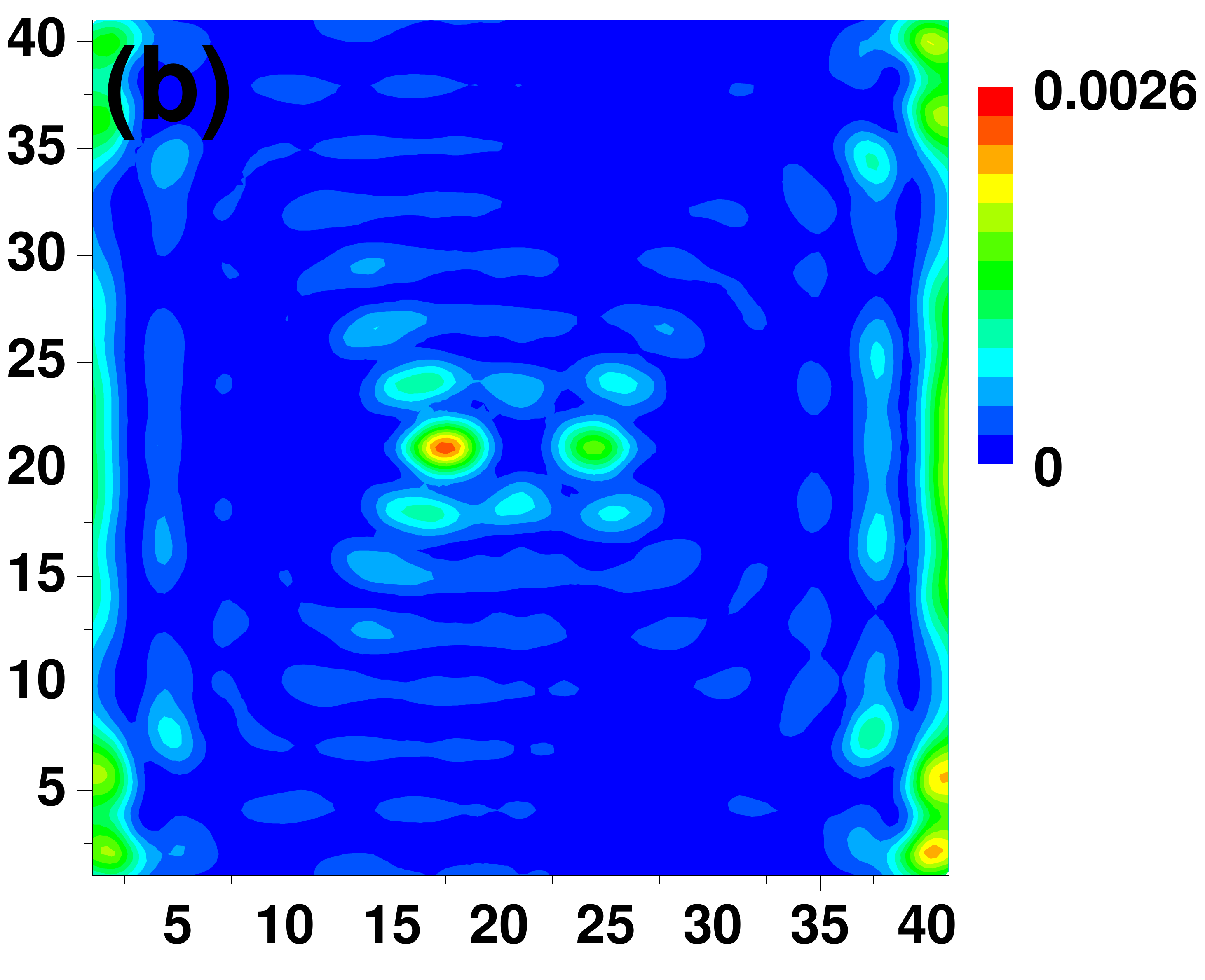} \\
\end{tabular}
\caption{(color online). The probability distribution of
quasiparticle for the Dresselhaus NCS plotted on the $41\times41$
square lattice. The parameters are $t=1$, $\beta=1$, $\Delta_{s}=1$.
The chemical potential and in-plane magnetic field are (a) $\mu=-4,
V^{2}=5$ and (b) $\mu=0, V^{2}=9$. }\label{fig3}
\end{figure}

To further study the Majorana fermions in the Dresselhaus NCS, we
consider the zero energy vortex core states by solving the BdG
equation for the superconducting order parameter of a single vortex
$\Delta(r,\theta)=\Delta\exp(i\theta)$ \cite{PhysRevB.79.094504}. To
do this, the $s$-wave superconducting term in the Hamiltonian Eq.
(\ref{eq1}) is modified to be position-dependent,
$H_{s}=\sum_{i}(\Delta
e^{i\theta_{i}}c_{i\uparrow}^{\dag}c_{i\downarrow}^{\dag}+\text{H.c.})$.
We numerically solve the Schr\"odinger equation $H\Psi=E\Psi$ for
the Hamiltonian in Eq. (\ref{eq1}), where
$\Psi=(u_{\uparrow},u_{\downarrow},v_{\uparrow},v_{\downarrow})^{T}$.
At zero energy we have $u_{\uparrow}=v_{\uparrow}^{*}$ and
$u_{\downarrow}=v_{\downarrow}^{*}$ as the particle-hole symmetry in
the Dresselhaus NCS, then the Bogoliubov quasiparticle operator,
$\gamma^{\dag}(E)=\sum_{i}(u_{i\uparrow}c_{i\uparrow}^{\dag}+u_{i\downarrow}c_{i\downarrow}^{\dag}+v_{i\uparrow}c_{i\uparrow}+v_{i\downarrow}c_{i\downarrow})$
becomes Majorana operator $\gamma^{\dag}(0)=\gamma(0)$. Therefore,
below we only consider the zero energy vortex core states for
discussing the MFs in the vortex core. Let's set the $x$ and $y$
directions to be open boundary, then we solve the BdG equations
numerically and calculate the density profile of quasiparticle for
the zero energy vortex core states. Previously, we have shown in the
Fig. (\ref{fig2}) that there is a novel semimetal phase in the
Dresselhaus NCS where the zero energy flat ABSs host MFs. Here we
shall ascertain whether there exist zero energy vortex core states
hosting MFs in this semimetal phase. The density profiles of
quasiparticle of the zero energy vortex core states are shown in the
Fig. (\ref{fig3}a) and (\ref{fig3}b), which correspond to the region
A and E in the phase diagram of the Fig. (\ref{fig1}b),
respectively. The numerical results of the energy for the zero
energy vortex core states are $E=2.54\times10^{-3}$ for Fig.
(\ref{fig3}a) and $E=6.68\times10^{-3}$ for Fig. (\ref{fig3}b),
respectively. For the choice of parameters in our simulations, the
order of magnitude of $\Delta_{s}^{2}/E_{F}$ is $10^{-1}$. Thus, the
numerical results have much smaller energy than the Caroli-de
Gennes-Matricon (CdGM) mode \cite{Caroli1964307}. It is clear to see
that there are zero energy states in the vortex core from the Fig.
(\ref{fig3}), therefore, the MFs exist in the vortex core of the
Dresselhaus NCS.

\section{summary}\label{summary}

In summary, we have investigated the topological phase and Majorana
fermion in the $s$-wave Dresselhaus (110) SO coupled NCS. We find
that there is a gapless region appearing in the phase diagram of the
Dresselhaus NCS. We observe that there exist flat Andreev bound
states which host Majorana fermions in the gapless region. The
chemical compound InSb has the largest Dresselhaus (110) SO coupling
\cite{PhysRevB.82.155327,PhysRevB.81.125318} which makes it
promising to observe the MFs at the edge or in the vortex core of
the system. We can fabricate an InSb (110) quantum well, in contact
with an $s$-wave superconducting aluminum and couple it to an
in-plane magnetic field. Thus we can apply the tunneling conductance
measurements to detect the zero-bias conductance peak of the system.

\begin{acknowledgments}
This work is partly supported by National Research Foundation and
Ministry of Education, Singapore (Grant No. WBS: R-710-000-008-271)
\end{acknowledgments}



%

\end{document}